\documentclass[aps,prl,twocolumn,showpacs]{revtex4}
\usepackage{amssymb}

\usepackage{graphicx,amsmath}

\newcommand{\bea}{\begin{eqnarray}}
\newcommand{\eea}{\end{eqnarray}}
\newcommand{\beq}{\begin{equation}}
\newcommand{\eeq}{\end{equation}}
\newcommand{\benu}{\begin{enumerate}}
\newcommand{\enu}{\end{enumerate}}

\begin{document}
\title{Kondo Breakdown and Hybridization Fluctuations in the
Kondo-Heisenberg Lattice}
\date{\today}
\author{I. Paul$^{1,2}$, C. P\'epin$^1$ and M. R. Norman$^2$}
\affiliation{$^1$SPhT, CEA-Saclay, L'Orme des Merisiers, 91191
Gif-sur-Yvette, France\\
$^2$Materials Science Division, Argonne National Laboratory, Argonne, IL
60439}

\begin{abstract}
 We study the deconfined quantum critical point of the
 Kondo-Heisenberg lattice in three dimensions using a fermionic representation
for  the localized spins. The mean-field phase
diagram exhibits a zero temperature quantum critical point separating a spin
liquid phase where the hybridization vanishes and a Kondo phase
where it does not.
Two solutions can be stabilized in the Kondo phase,
namely a uniform hybridization
when the band masses of the conduction electrons and the spinons
have the same sign, and a modulated one when they
have opposite sign. For the uniform case,
we show that above a very small  temperature    scale,
the critical fluctuations associated with the vanishing hybridization
have dynamical exponent $z=3$, giving rise to a resistivity that has
a $T \log T$ behavior.  We also find that the
specific heat coefficient diverges logarithmically in temperature,
as observed in a number of heavy fermion metals.
\end{abstract}

\maketitle
A large number of experiments have been performed on
  metallic
 heavy fermion compounds close to a  zero temperature phase transition
 (a quantum critical point (QCP))
driven by applied magnetic field, chemical doping or
pressure \cite{stewart}. In the quantum critical regime,  the
thermodynamics and transport properties are very unusual, violating
the predictions of the
Landau Fermi liquid theory of metals. The resistivity is quasi-linear
in temperature over several decades, and in many cases
the specific heat coefficient diverges logarithmically
as the temperature is decreased.
These unusual observations have motivated many
theoretical studies that have attempted to capture these effects.
Most theories~\cite{hertz,millis,moriya,rosch}
are based on the assumption that at the
QCP,  the Fermi liquid is destabilized by spin density wave formation,
and therefore the critical fluctuations are
magnetic in nature.
In $d=3$, these
theories fail to capture simultaneously the linear
temperature dependence of the
resistivity and the divergence of the specific heat coefficient at low
temperatures~\cite{review-piers}.
More recently the problem has been approached from another
perspective which takes the point of view that
at the QCP,
magnetic fluctuations  suppress the formation of the heavy Fermi liquid,
driving the effective Kondo temperature of the
lattice ($T_K$) to zero~\cite{review-piers,qimiao,senthil,schofield}.
In this picture, the QCP
is  a bi-critical point where the metal experiences
fluctuations due to the vanishing energy scale $T_K$ as well as the paramagnons.
One feature that distinguishes between these two classes of theories is that
the first predicts the Fermi volume to change smoothly across
the QCP, while the second predicts an abrupt change~\cite{review-piers}.


In this Letter we explore the possibility that
in the quantum critical regime,
the unusual
behavior in thermodynamics and transport is due to
critical fluctuations, but of a non-magnetic order parameter
associated with the vanishing energy scale $T_K$,
and not due to paramagnons.
 The order parameter we
advocate is the field $\sigma$ associated with the hybridization between
the localized
spins and the conduction electrons~\cite{read,millis-lee}.
At the QCP,  the effective
Kondo temperature for the lattice goes to zero, leading
to a ``Kondo breakdown'' of the heavy Fermi liquid.
The critical fluctuations of $\sigma$ are gapless excitations,
and we study how these fluctuations influence
the properties of the metal using
 the formalism of the large $N$ Kondo-Heisenberg model.

 Beyond the mean-field level, the
Kondo-Heisenberg model can be treated as a lattice gauge theory.
 Senthil {\it et.~al.}~\cite{senthil} have examined the effect of the gauge fluctuations in
 this model, while
Coleman {\it et.~al.}~\cite{schofield} studied the zero temperature transport
 anomalies.  In our work, we find a number of novel effects associated with
 the fluctuations of the $\sigma$ field which were not discovered in these earlier studies.

At the Kondo breakdown  QCP where
$\langle \sigma \rangle =0$, we observe two new phenomena:
(1) $\sigma$ can order at a finite wavevector
leading to spatial modulations of the Kondo hybridization
analogous to the LOFF state  of superconductivity \cite{fflo};
(2) the presence of multiple energy scales, spread over a very large
range in energy.  The lowest scale is extremely small (of order 1 mK),
above which, up to an ultraviolet cutoff of order the single ion Kondo temperature,
the critical fluctuations of $\sigma$ exhibit a
dynamical exponent $z=3$. This gives rise to a marginal Fermi liquid like
behavior for the conduction electrons in $d=3$, with
a resistivity that goes as $T \log T$.  A logarithmic dependence is also found
for the specific heat coefficient from both the gauge
and $\sigma$ fluctuations.


Our starting point is the large $N$ formulation of the
Kondo-Heisenberg model, where $N$ denotes the enlarged spin
symmetry group $SU(N)$. It describes a broad band of  conduction
electrons interacting with localized spins through
anti-ferromagnetic Kondo coupling $J_K > 0$. The localized spins
interact with each other via nearest neighbor exchange $J_H > 0$.
 We work with a representation of the localized spins in terms of
Abrikosov pseudo fermions
$ {\vec S_i} = \sum_{\alpha \beta}
 f^\dagger_{i, \alpha} {\vec \sigma}_{\alpha \beta} f_{i, \beta}$, where
 $(\alpha, \beta) = (1, N)$,
with the constraint of $n_f= N/2$ spinons per site $i$. The
interactions which are quartic fermionic terms can be decoupled
using Hubbard-Stratonovich fields $\varphi_{ij} \rightarrow
\sum_{\alpha} f^\dagger_{i \alpha} f_{j \alpha} $ for the
Heisenberg exchange, and $ \sigma_i^\dagger \rightarrow
\sum_\alpha f^\dagger_{i \alpha} c_{i \alpha} $ for the Kondo
interaction.
Following Ref.~\onlinecite{senthil} we assume that in $d=3$,
$\varphi$ condenses in a uniform spin liquid phase that
gives dispersion to the spinons, which is an essential ingredient for
Kondo breakdown to occur (physically we interpret the uniform spin
liquid as a mean field description of the short range magnetic
correlations that persist when a magnetic ground state
is destroyed by quantum fluctuations).
This
gives the Lagrangian \bea \label{eqn2} \mathcal{L}  \! \! &=& \!
\! \! \! \sum_{\langle ij \rangle \alpha} \! \left[ c^{\dagger}_{i
\alpha} \! \left(  \partial_{\tau} + t_{ij} \right) c_{j \alpha} +
f^\dagger_{i   \alpha} \! \left ( \partial_{\tau} + \varphi_0 e^{
i a_{ij}} + \lambda_i \delta_{ij} \right ) f_{j \alpha} \right]
\nonumber \\
&-&
\frac{N}{2} \sum_i \lambda_i + \frac{N}{J_K} \sum_i \sigma^{\dagger}_i \sigma_i
+ \frac{N\varphi_o^2}{J_H} \nonumber \\
&+&
 \sum_{i \alpha} \left( c^\dagger_{i \alpha} f_{i \alpha} \sigma_i
+ \rm{h.c.} \right) \eea ($V$, the volume of the system, is set to
1). The above Lagrangian has a local $U(1)$ gauge
invariance~\cite{ioffe}. The Lagrange multiplier $\lambda_i$
(scalar potential) enforces the constraint $n_f = N/2$ per site.
 Given a state which satisfies the above constraint, a
single hop of a spinon will violate it. Consequently only
simultaneous opposite hops of spinons between two neighboring
sites are physically allowed. This implies that the local spinon
current operator $J_{f i} =0$ at each site. The gauge fields
$a_{ij}$ (vector potential) associated with the phase of
$\varphi_{ij}$ ensure that this condition is satisfied.

There are two important parameters in
Eq.~\ref{eqn2}. First, $\alpha = \varphi_0/D$, which is the
ratio of the spinon to the conduction electron bandwidth $D$ (note from Eq.~\ref{eqn2}
that for $\sigma$=0, $\phi_0=J_H$).
Second, while the spinon band is half-filled due to the constraint
(henceforth we assume $N=2$),
the conduction band filling is generic. Without any loss of generality we
take the conduction band to be less than half filled. This implies that the Fermi
wavevector of the spinon band, $k_{F0}$, is different from that of the conduction
band, $k_F$. We denote the mismatch by $q^{\ast} = k_{F0} - k_F$. In the following
we take $\alpha$ and $(q^{\ast}/k_F)$ to be small.
We identify $\alpha D$ with the single ion Kondo scale ($T_K^0 =D e^{-1/\rho_0 J_K}$)
which is typically $10$K in
heavy fermions. Assuming $D \sim 10^4$K, we get $\alpha \sim 10^{-3}$.

At the mean-field level, the parameters $\varphi_0$,
 $\langle \lambda_i \rangle$ and $\langle \sigma_i \rangle$
 are determined by minimizing the free energy, $F$. The mean-field phase transition between
 the spin liquid state, $\langle \sigma_i \rangle = 0$, and the heavy Fermi liquid state
 with a lattice Kondo temperature $T_K \approx \pi \rho_0 \langle \sigma_i \rangle^2$,
 occurs when
 \beq \label{eqn4}
\frac{\partial^2 F}{\partial \left| \sigma_q \right|^2 } = \frac{1}{J_K} +
\Pi_{fc}(q,0) = 0
\eeq
where $\Pi_{fc} (q, 0)$ is the static electron-spinon (fc) polarization. We solve
this equation for two different situations, the result of which is depicted
in Fig.~1. (i) e-e case, where both the bands are taken to be electron-like.
Linearizing the fermionic dispersions we have
$\epsilon_k = v_F (k - k_F)$ for the conduction electrons, and
$\epsilon^0_k = \alpha v_F (k - k_F - q^{\ast})$ for the spinons (where $k=|{\bf k}|$). For linearized
dispersions, $\Pi_{fc}(q,0)$ turns out to be $q$-independent. Inclusion of the curvature
stabilizes a second order phase transition around $q = 0$, the polarization taking the form
$\Pi_{fc}(q,0)= \frac{\rho_0}{1-\alpha}(\ln\alpha + \frac{1-\alpha^2+2\alpha\ln\alpha}{4(1-\alpha)^2}
\frac{q^2}{k_F^2})$ where
$\rho_0=1/D$ is the conduction electron density of states at the Fermi energy. In this case the
$T = 0$ phase transition occurs at a critical Kondo coupling of
$J_K^c = 1/(\rho_0 \ln (1/ \rho_0J_H))$ \cite{tkfoot}. (ii) e-h case, where the conduction band is taken
to be electron-like as before, while the spinon band
is hole-like with a linearized dispersion
$\epsilon^0_k = -\alpha v_F (k - k_F - q^{\ast})$. In this case we find
$\Pi_{fc}(q,0) = \frac{\rho_0}{1+\alpha}
(\ln\frac{\alpha v_F^2 |q^{*2}-q^2|}{D^2(1+\alpha)^2} - 2 +
\frac{q^*}{q}\ln\frac{q^*+q}{|q^*-q|})$, which has a minimum at $q = 1.2 q^{\ast}$
independent of $\alpha$.
In this state $T_K$ is modulated, with nodes in space where $T_K$ vanishes.
This solution is similar to the spin density wave instability
encountered in chromium \cite{rice}  and in the LOFF state of
superconductivity \cite{fflo,foot2}. Fig.~1a shows that for parabolic bands
the minimum  of the effective potential
is lower in the e-h case than in the e-e case. Thus, for parabolic bands,
the  modulated solution is more stable (Fig.~1b).
However the
question of which solution is realized in real compounds will be
material dependent.


\begin{figure}

\includegraphics[width=3.4in]{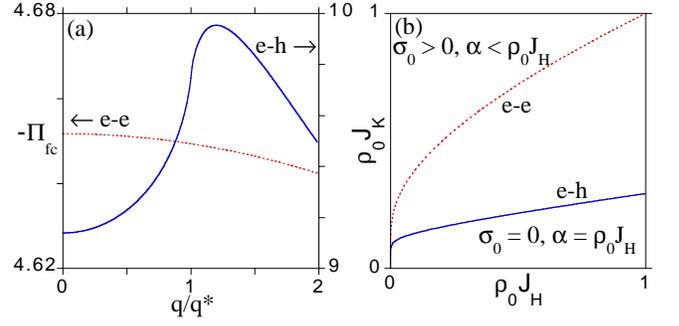}
\caption{(a) q dependence of $\Pi_{fc}$ ($\alpha$=0.01
and $q^*/k_F$=0.1).  Note differing scales for the e-e and e-h cases.
(b) Quantum critical point as a function of $J_K$ and $J_H$ for the e-e (q=0)
 and e-h
(q=1.2q*) cases.}
\label{fig1}
\end{figure}

We now turn to the fluctuations around the mean-field solution.
We  present our results for the simpler e-e case, leaving
the more complex e-h case for a later paper. In the quantum critical regime there
are two important types of gapless fluctuations, namely  the gauge fluctuations
associated with $a_{ij}$ and the critical fluctuations of the Kondo bosons $\sigma$.

The gauge fluctuations of this theory have been studied earlier by
Senthil \emph{et. al.}~\cite{senthil}. Here we summarize the salient points to put
our work in perspective.
It is convenient to work in the Coulomb gauge ${\vec \nabla} \cdot {\vec a} =0$,
 where the vector gauge fields
 $a_{\mu}$ ($\mu = x, y,z)$ are purely transverse \cite{ioffe}. The fluctuations
of $\lambda$ decouple from those of $a_{\mu}$, and give rise to a
screened Coulomb interaction. As such, they are massive and can be
neglected. The gauge fields $a_{\mu}$, which act as vectorial
Lagrange multipliers to ensure that the local spinon current $J_{f
i} = 0$, do not have any intrinsic dynamics of their own. Their
dynamics is generated via coupling with the spinon band, and
therefore they are over-damped. For frequencies smaller than the
spinon bandwidth $ \alpha D$, the transverse gauge field
propagator $D_{\mu \nu} ({ \bf x}, \tau ) = \langle T_{\tau} \left
[a_\mu ({\bf x}, \tau ) a_\nu (0,0) \right ] \rangle$ has the
standard form $D_{\mu \nu} ( q, i \Omega_n) = ( \delta_{\mu \nu} -
q_\mu q_\nu/ q^2 ) \Pi^{-1} (q, i \Omega_n)$, with $\Pi (q, i
\Omega_n) \propto [ (q/2k_{F0})^2 + |\Omega_n|/(\alpha v_F q)]$.
This is the typical form for excitations with dynamical exponent
$z=3$, which  in $d=3$ are known \cite{reizer} to give a contribution to the
specific heat coefficient $\gamma \equiv - \partial^2F/\partial
T^2 \propto \ln(\alpha D/T)$ and to
the static spin susceptibility $\delta \chi_s \propto T^2\ln(\alpha D/T)$.
Finally, when the compact nature of the $U(1)$ gauge
group on the lattice is taken into account, the gauge fluctuations convert the finite
temperature mean-field transition line into a crossover line~\cite{senthil,nagaosa-lee}.

\begin{figure}
\includegraphics[width=3.4in]{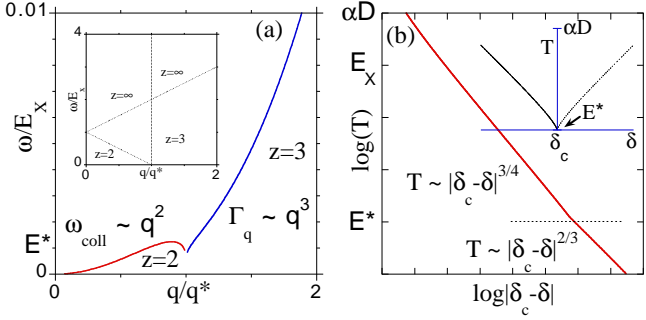}
\caption{
(a) Structure of $D_{\sigma}$ for the e-e case ($\alpha << 1$) at the QCP for positive frequencies.
$z$ is the dynamical exponent in the various regions, which are delineated by the  dashed
lines, equal to $\alpha v_F (q^{\ast} \pm q)$ and $v_F (q^{\ast} \pm q$).  $\omega_{coll}$
is the zero of $D_{\sigma}^{-1}$ in the z=2 regime (a propagating mode),
and $\Gamma_q$ is the dispersion of the damped mode in the z=3 regime
(maximum of the imaginary part of $D_{\sigma}$).
Note  the presence of energy scales $E^{\ast} \sim 10^{-4} \alpha D$ and
 $E_X= \alpha v_F q^{\ast} \sim 10^{-1} \alpha D$.
(b) Phase diagram for the e-e case ($\alpha << 1$).  Note the crossover from z=2 to z=3
behavior at $E^{\ast}$ (dotted line).  The inset shows the phase diagram on a linear scale.  The solid
line is the crossover line in the Kondo phase, the dashed line the crossover
line in the spin liquid phase.}
\label{fig2}
\end{figure}

In the quantum critical regime, the fluctuations of the complex order parameter
fields ($\sigma^{\dagger}_i, \sigma_i$) are massless as well (ignoring for the moment the
$T$-dependent mass generated by the quartic $|\sigma|^4$ coupling). The propagator
for these fluctuations is defined as $D_{\sigma} (x, \tau) =
\langle T_{\tau} \left [ \sigma^{\dagger} (x, \tau ) \sigma ( 0,0 ) \right ]
\rangle$
with
$D^{-1}_\sigma ( q, i \Omega_n )=  1/ J_K + \Pi_{fc}(q,0)+\Delta\Pi_{fc} ( q, i \Omega_n ) $ where
\beq
\Delta\Pi_{fc} ( q, i \Omega_n ) = \sum_{\pm} \frac{  \rho_0 [
\mp X_{1 \pm} \ln ( X_{1 \pm})
 \pm X_{2 \pm} \ln ( X_{2 \pm} )] }{ 2 \alpha v_F q (1 - \alpha ) }   \eeq
with $X_{1 \pm } =
-  \alpha i \Omega_n  \pm \alpha v_F q - \alpha v_F q^*$, $X_{2 \pm} = - i \Omega_n
\pm \alpha v_F q - \alpha v_F q^*$, and $\Pi_{fc}(0,0)=\rho_0\ln(\alpha)/(1-\alpha)$
($\Delta\Pi_{fc}$ is the dynamical part of the fc polarization).
The
different regimes of $D_{\sigma}(q, i \Omega_n)$ with their
associated dynamical exponents $z$ are summarized in Fig.~2a.  At
high energies, one finds $z=\infty$ behavior consistent with
quasi-local behavior.  But we find physical properties are
dominated by the $z=3$ and $z=2$ regimes. These two regimes can be
understood as follows. Due to the mismatch between the two Fermi
surfaces, a minimum momentum is necessary to excite inter-band
(fc) particle-hole pairs. As a result for $\Omega < \alpha v_F
(q^{\ast}-q)$, excitations of $\sigma$ do not decay into
particle-hole pairs but propagate ballistically with
$D^{-1}_{\sigma}(q, i \Omega_n) \approx \rho_0[q^2/(4k_F^2) - i
\Omega_n/E_X]$, i.e., $z=2$ (where $E_X=\alpha v_F q^{\ast}$).
This behavior is cutoff for frequencies $\Omega > E^{\ast} $ with
\beq \begin{array}{ll}\label{estar} E^{\ast}= c \alpha D
(q^{\ast}/k_F)^3  \; \; & \mbox{and} \;  c \sim 10^{-1} \ ,
\end{array} \eeq
above which the dynamical exponent $z$ changes from 2 to 3.
For most of the phase space,  the spectrum for the
fluctuations of $\sigma$ lie within the inter-band particle-hole
continuum, making their dynamics over-damped with $D^{-1}_{\sigma}(q, i
\Omega_n) \approx \rho_0[q^2/(4k_F^2) +
\pi|\Omega_n|/(2\alpha v_F q)]$, i.e., $z=3$. The energy scale in
the z=3 regime is $\Gamma_q/E_X  \approx q^3/(2\pi k_F^2q^{\ast})$
which has an infrared cutoff at $E^{\ast}$ because of the
mismatch vector $q^{\ast}$. The ultraviolet cutoff scale for the
$z=3$ regime is $\alpha v_F (q+q^{\ast})$ which is of order
$\alpha D$ for $q \sim k_F$.  The energies $E^{\ast}$ (infrared)
and $\alpha D$ (ultraviolet) appear as crossover scales for any
physical property that is affected by the excitations of $\sigma$.
For a one impurity Kondo scale $\alpha
D \sim 10 K$,  and  $q^{\ast}/k_F \sim 10^{-1}$ \cite{foot3}, we estimate $E^{\ast}
\sim 1 mK$. $E^{\ast}$ is therefore a very small energy
scale which is essentially unobservable.

The crossover lines in $T$ that define the quantum critical region are symmetric around the
QCP $\delta=\delta_c$, where $\delta= 1/(\rho_0 J_K)$ (Fig.~2b). These are determined
by the $T$-dependent mass generated by the quartic $|\sigma|^4$ coupling. For
$T < E^{\ast}$, we find that the leading contribution is from the $z=2$ regime
(proportional to $T^{3/2}$ for $d=3$), so that the crossover temperature
$T \propto |\delta-\delta_c|^{2/3}$, while for $T > E^{\ast}$ the $z=3$ regime dominates giving
a crossover temperature $T \propto |\delta - \delta_c|^{3/4}$ for $d=3$.

Next we examine the contribution to the free energy from the
fluctuations of $\sigma$. We find that the leading contribution is
entirely due to the $z=3$ regime since it comes from a much larger
phase space volume (the $z=2$ contribution is similar to that of a
gapless magnon mode). For $E^{\ast} < T < \alpha D$, we find
$F(T) \sim \int n_B Im \ln (D_{\sigma}^{-1}) \sim - k_F^3T^2/(9\alpha D) \ln (\alpha D/T)$,
which is a typical
result for $z=3$ excitations. This implies a contribution to the
specific heat coefficient $\gamma \sim 2k_F^3/(9\alpha D)\ln(\alpha D/T)$, which
adds to a similar contribution from the transverse gauge
fluctuations. For $T < E^{\ast}$, the infrared cutoff sets in, and
the specific heat coefficient from the $\sigma$ fluctuations
saturates.  This regime is then dominated by the logarithmic
contribution from the transverse gauge fields \cite{reizer}.

We now calculate the self energy of the conduction electrons due to the hybridization
fluctuations ($c \rightleftharpoons f + \sigma$). This is defined as
$\Sigma_c (k, i \omega_n) = T \sum_{\omega_n, q}
G_f(k+q, i\omega_n + i\Omega_n) D_{\sigma}(q, i\Omega_n)$, where
 $G_f^{-1}(k, i \omega_n) = (i \omega_n - \epsilon^0_k)$ is
the inverse propagator of the spinons.
As in the case of the free energy, we find that the leading
contribution is due to the $z=3$ regime of $D_{\sigma}(q, i\Omega_n)$. At $T=0$ and
for $E^{\ast} < \omega < \alpha D$, we find
$
Im\Sigma_c(k_F,\omega) \sim k_F^2/(6\pi\alpha v_F \rho_0)\omega.
$
The temperature dependence of
${\rm Im} \Sigma_c(k_F, \omega=0, T)$ involves a frequency integral weighted
by the factor $n_B + n_F
= 1/\sinh(\Omega/T)$. This makes the integral
logarithmically divergent in the infrared, which is cutoff by $E^{\ast}$. For
$E^{\ast} < T < \alpha D$ we find
\beq
\label{eq:Sigma}
{\rm Im} \Sigma_c(k_F, \omega =0, T) \sim k_F^2/(6\pi\alpha v_F \rho_0)
T \ln (2T/E^{\ast}).
\eeq
For $\omega, T < E^{\ast}$, the self energy is Fermi liquid like.

We turn to the $T$-dependence of the static spin
susceptibility, $\chi_s(T)$.
At the mean field level, we find what is usual for band fermions, namely a
constant part plus a $T^2$ term.
To calculate the correction beyond mean field ($\delta \chi_s$) due to the Kondo bosons, we note
that in a magnetic field $B$, there is an additional $(B/(\alpha D))^2$ contribution to
$D_{\sigma}^{-1}(q, i\Omega_n)$. This gives
$\delta \chi_s (T) \propto T^{4/3}$ for $E^{\ast} < T < \alpha D$,
and a $T^2$
dependence below $E^{\ast}$ (so below $E^{\ast}$ the $T^2\ln(T)$
contribution due to the gauge bosons dominates).

Finally we discuss the temperature dependence of the resistivity,
$\rho$, that is obtained in the quantum critical regime.
Eq.~\ref{eq:Sigma} gives the $T$-dependence of the inverse
lifetime $\tau_{c}^{-1} \propto {\rm Im} \Sigma_c(T)$ of the
conduction electrons. For one band models experiencing $q \simeq
0$ scattering, this lifetime cannot be associated with the
transport lifetime, $\tau_{\rm tr}$, because the leading
contribution to the self energy comes from forward scattering
processes which are not effective in relaxing the current.
Consequently, when vertex corrections are taken into account,
$\tau_{\rm tr}^{-1}$ acquires an additional temperature dependence
proportional to $q^2 \sim T^{2/z}$.
However,  our model consists of two bands, one of
light particles (the conduction electrons)  which scatter from
very heavy particles (the spinons) \cite{ziman}.
As such,
the charge neutral spinons act as an effective bath for the
relaxation of the conduction electron current (the other charge
carrying modes, the complex $\sigma$ bosons, have over-damped
dynamics, and therefore the current is mostly carried by the
conduction electrons).
The first non-zero vertex correction involves two
$\sigma$ boson exchange processes. Such a correction is small by a
factor of $\alpha$. Therefore $\tau_{\rm tr}$ can be identified
with $\tau_{c}$, and for $E^{\ast} < T < \alpha D$ we find \beq
\delta \rho (T) \equiv \rho (T) - \rho(0) \propto T \ln
(2T/E^{\ast}). \eeq For $T < E^{\ast}$, $\delta \rho (T) \propto
T^2$, but $E^{\ast}$ is
extremely small ($\sim$ 1 mK). Thus, the Kondo-Heisenberg model
captures one of the most mysterious features of quantum
criticality in heavy fermion compounds, namely the quasi-linear
resistivity observed for most compounds over a large temperature
range.

In conclusion, we studied the Kondo breakdown QCP of the Kondo-Heisenberg model
in $d=3$. Over a large temperature range, we find that
the critical fluctuations have a dynamical exponent $z=3$, giving rise to marginal Fermi liquid
behavior for the conduction electrons. The specific heat coefficient has
a $\log (1/T)$ dependence, while the resistivity has a $T \log T$ behavior.  The Kondo-Heisenberg
model is characterized by multiple energy scales, and as such shows great promise in
explaining the various subtleties associated with heavy fermion quantum  critical behavior.

We thank the hospitality of the KITP where this work was
initiated. We also acknowledge  J. Schmalian, P. Sharma, and A.
Chubukov for extensive discussions. This work was supported by the
U.S. Dept.~of Energy, Office of Science, under Contract
No.~W-31-109-ENG-38 and in part by the National Science Foundation
under Grant No.~PHY99-07949.


\begin{thebibliography}{99}
\bibitem{stewart} G. Stewart, Rev. Mod. Phys. {\bf 73}, 797 (2001).
\bibitem{hertz} J. A. Hertz, Phys. Rev. B {\bf 14}, 1165 (1976).
\bibitem{millis} A. J. Millis, Phys. Rev. B {\bf 48}, 7183 (1993).
\bibitem{moriya} T. Moriya and T. Takimoto, J. Phys. Soc.  Japan {\bf 64}, 960 (1995).
\bibitem{rosch} A.  Rosch, A. Schroder, O. Stockert and H. von Lohneysen, Phys. Rev. Lett. {\bf 79},
159 (1997); A. Rosch, {\it ibid} {\bf 82}, 4280 (1999).
\bibitem{review-piers} P. Coleman, C. Pepin, Q. Si and R. Ramazashvili, J. Phys.: Condens. Matter
{\bf 13}, R723 (2001).
\bibitem{qimiao} Q. Si, S. Rabello, K. Ingersent and J. L. Smith, Nature {\bf 413}, 804 (2001);
D. R. Grempel and Q. Si,
Phys. Rev. Lett. {\bf 91}, 026401 (2003);
P. Sun and G. Kotliar, Phys. Rev. Lett. {\bf 91},
037209 (2003).
\bibitem{senthil} T. Senthil, S. Sachdev and M. Vojta, Phys. Rev. Lett. {\bf 90}, 216403 (2003);
T. Senthil, M. Vojta and S. Sachdev, Phys. Rev. B {\bf 69}, 035111 (2004).
\bibitem{schofield} P. Coleman, J. B. Marston and A. J. Schofield, Phys. Rev. B {\bf 72}, 245111 (2005).
\bibitem{read} N. Read and D. M. Newns, J. Phys. C {\bf
16}, 3273 (1983); N. Read, J. Phys. C {\bf 18}, 2651 (1985).
\bibitem{millis-lee} A. J. Millis and P. A Lee, Phys. Rev. B {\bf 35}, 3394 (1987).
\bibitem{fflo} P. Fulde and R. A. Ferrell, Phys. Rev. {\bf 135}, A550 (1964);
A. I. Larkin and Y. N. Ovchinnikov, Sov. Phys. JETP {\bf 20}, 762 (1965).
\bibitem{ioffe} L. B. Ioffe and A. I. Larkin, Phys. Rev. B {\bf 39}, 8988 (1989);
P. A. Lee and N. Nagaosa, Phys. Rev. B {\bf 46}, 5621 (1992).
\bibitem{tkfoot} That is, $\alpha D = D e^{-1/\rho_0J_K^c} = T_K^0$.
\bibitem{rice} T. M. Rice, Phys. Rev. B {\bf 2}, 3619 (1970).
\bibitem{foot2} For spherical Fermi surfaces, as many ordering
wave vectors as allowed by lattice symmetry will condense,
each with a modulus of $q$.
\bibitem{reizer} T. Holstein, R. E. Norton  and P. Pincus, Phys. Rev. B {\bf 8}, 2649 (1973);
M. Yu. Reizer, {\it ibid} {\bf 40}, 11571 (1989).
\bibitem{nagaosa-lee} N. Nagaosa and P. A. Lee, Phys. Rev. B {\bf 61}, 9166 (2000).
\bibitem{foot3} Heavy fermion metals have complex Fermi surfaces where both conduction
and f surfaces are large.  As a consequence, large values of $q^{\ast}$ are unlikely.
Very small values of $q^{\ast}$ are possible due to degeneracies.  Within our simple model, a
value of $q^{\ast}/k_F \sim 0.1$ is a reasonable estimate.
\bibitem{ziman} In a simple Boltzman approximation for f-c scattering, the factor weighting
the transport
would be $1-\alpha \cos(\theta) \simeq 1$, see J. M. Ziman, {\it Electrons and Phonons}
(Oxford Univ. Pr., London, 1960), p. 376-377.


\end{thebibliography}
\end{document}